\documentclass[10pt,twocolumn,letterpaper]{article}

\usepackage[pagenumbers]{iccv}
\definecolor{iccvblue}{rgb}{0.21,0.49,0.74}
\usepackage[pagebackref,breaklinks,colorlinks,allcolors=iccvblue]{hyperref}

\def\paperID{9769} 
\def\confName{ICCV}
\def\confYear{2025}

\title{ReCoM: Realistic Co-Speech Motion Generation with Recurrent Embedded Transformer}

\author{
Yong Xie$^{1}$ \quad Yunlian Sun$^1$ \quad Hongwen Zhang$^{2}$ \quad Yebin Liu$^{3}$  \quad Jinhui Tang$^{1}$ \\
$^1$Nanjing University of Science and Technology \quad $^2$Beijing Normal University  \quad $^3$Tsinghua University
}

\begin{document}
\maketitle
\begin{abstract}
We present ReCoM, an efficient framework for generating high-fidelity and generalizable human body motions synchronized with speech. The core innovation lies in the Recurrent Embedded Transformer (RET), which integrates Dynamic Embedding Regularization (DER) into a Vision Transformer (ViT) core architecture to explicitly model co-speech motion dynamics. This architecture enables joint spatial-temporal dependency modeling, thereby enhancing gesture naturalness and fidelity through coherent motion synthesis. To enhance model robustness, we incorporate the proposed DER strategy, which equips the model with dual capabilities of noise resistance and cross-domain generalization, thereby improving the naturalness and fluency of zero-shot motion generation for unseen speech inputs.
To mitigate inherent limitations of autoregressive inference, including error accumulation and limited self-correction, we propose an iterative reconstruction inference (IRI) strategy. IRI refines motion sequences via cyclic pose reconstruction, driven by two key components: (1) classifier-free guidance improves distribution alignment between generated and real gestures without auxiliary supervision, and (2) a temporal smoothing process eliminates abrupt inter-frame transitions while ensuring kinematic continuity.
Extensive experiments on benchmark datasets validate ReCoM's effectiveness, achieving state-of-the-art performance across metrics. Notably, it reduces the Fréchet Gesture Distance (FGD) from 18.70 to 2.48, demonstrating an 86.7\% improvement in motion realism. Our project page is \url{https://yong-xie-xy.github.io/ReCoM/}.
\end{abstract}    
\section{Introduction}
\label{sec:intro}

\begin{figure*}
	\centerline{\includegraphics[width=\textwidth]{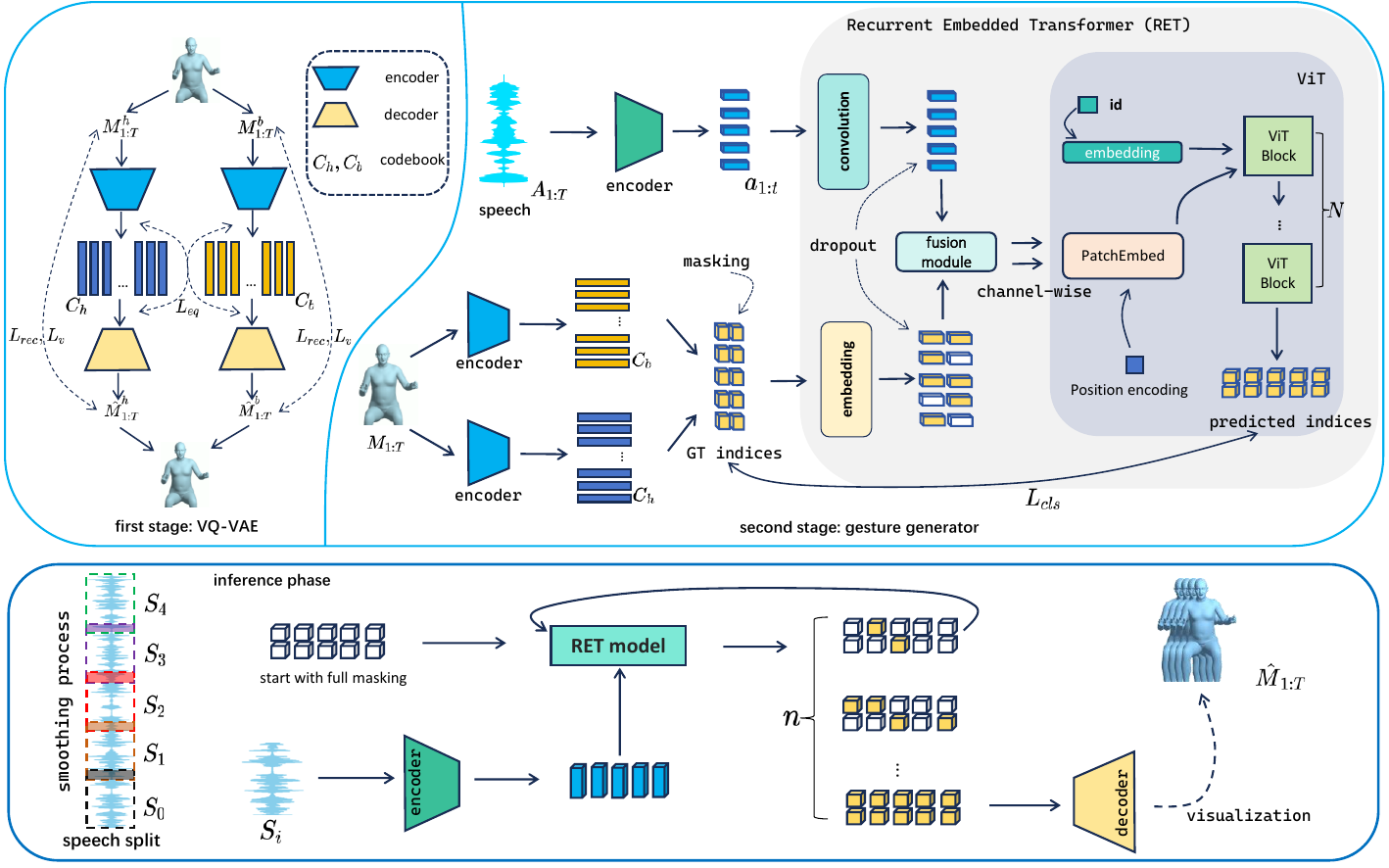}}
	\caption{The gesture training and inference pipeline of our work. Given an audio input, our method aims to produce a high-fidelity gesture. We use the loss function $L_{VQ}$ to optimize the compositional VQ-VAEs, enabling them to learn discrete gesture representation. We carefully employ effective data processing strategies to optimize the gesture generator, enabling it to obtain high-fidelity results. In the Inference phase, $S_{i}$ denotes the $i$-th speech segment. In this phase, we use IRI and a temporal smoothing process. The RET model is our gesture generator. Additionally, $n$ denotes the number of iterations, which is a variable value depending on the input. The novel inference strategy further enhances the model's performance in a non-autoregressive way. Most of the variables mentioned in the paper are introduced in \cref{sec:pipelineOverview}, while the remaining variables are mentioned in \cref{sec:FaceGenerator}, \cref{sec:GestureCodebook}, and \cref{sec:GestureGenerator}.}
	\label{fig:EDIT}
\end{figure*}

Human motion generation is a broad concept aimed at creating natural, realistic, and diverse human motions, including whole-body movements such as walking, jumping, dancing, etc. Recently, many excellent works have emerged, such as \cite{motionclip,omg,momask,t2m-gpt,fg-mdm,ListenAction}, all of which have made outstanding contributions to the field of human motion generation.

Co-speech gesture generation is an important subtask in the field of human motion generation. It involves the automatic generation of expressions and gestures that are seamlessly aligned with speech. And it primarily focuses on upper-body movements, including the motion of the body, hands, and face. This research area holds significant importance for human-computer interaction, digital entertainment, virtual reality, intelligent robots, and other domains. These gestures can help to enrich speech presentation, thus achieving a more natural and fascinating communication experience.

Currently, co-speech gesture generation technology primarily relies on deep learning models. Popular models include Generative Adversarial Networks (GANs) \cite{GANs}, Variational Autoencoders (VAEs) \cite{VAE}, Vector Quantized Variational Autoencoders (VQ-VAE) \cite{VQ-VAE}, and Diffusion Models \cite{DDPM,DDIM,LDM}. These models are capable of understanding the intricate interactions between speech signals and gestures, enabling them to generate gestures that are consistent with the speech content.

Presently, the field of gesture generation has drawn significant attention. Studies such as those in \cite{talkshow,emage} use VQ-VAE. \cite{SemanticGS} uses the residual VQ-VAE \cite{RQVAE}. Studies in \cite{beat,emage} manage to perform the step-by-step processing and transformation of data by means of cascaded architectures. The study in \cite{cocogesture} takes DiT \cite{DiT} as its main architecture. Furthermore, studies \cite{gesturediffuclip,diffgesture,diffsheg,amuse} adopt diffusion models as their principal architectures. Moreover, the research in \cite{ProbTalk} introduces product quantization to the VAE and enriches the representation of complex holistic motion.

Though achieving impressive results, these approaches face some challenges. One challenge is how to ensure that the model learns appropriate gesture features that meet the demand of fidelity and generalization. Another challenge is that large gesture datasets are difficult to acquire. Fortunately, many good pose estimation methods were proposed \cite{recoverysurvey,recovery1,recovery2,recovery3}, fulfilling the need for acquiring high-quality datasets. Hence, we put our emphasis on addressing the first challenge.

Our motivation for ReCoM mainly stems from two observations. Firstly, we noticed that previous methods don't adequately fit the dataset, meaning these models might lack sufficient learning ability. For instance, we observed that the Habibie \etal method \cite{Habibie} generates gestures with excessively large jitter amplitudes. TalkSHOW \cite{talkshow}, which occasionally produces jittery motions, frozen movements, and noticeable penetration in animation, results in low fidelity. Secondly, we observed that previous studies don't pay sufficient attention to generalization, meaning that their performance on out-of-domain datasets is poor. For instance, ProbTalk \cite {ProbTalk} demonstrates suboptimal performance on generalization datasets, and sometimes its visualized motions are overly slow. Given humans' strong perceptual sensitivity to unnatural human motions, this series of problems may lead to a poor user experience when applied.

In addition, the works mentioned above rarely focus on both temporal and spatial information simultaneously. Due to the aforementioned issues, we propose some corresponding improvement strategies. Our method is based on \cite{talkshow}, and our solution draws inspiration from \cite{ViT}, and \cite{momask}. Since handling spatial information presents more challenges, we focus on this issue in this work. Specifically, we leverage ViT to design our core architecture: Recurrent Embedded Transformer (RET). RET incorporates channel-wise operations that optimize it for spatial information processing. Other architecture choices include adopting VQ-VAE to learn two discrete gesture representations respectively for body and hand, and using an encoder-decoder architecture for face generation.

Our model's competitive results is evident from the comparisons detailed in \cref{sec:Comparisons}. We believe that the effectiveness of our method mainly stems from \textbf{three improvements}. The first improvement involves an innovatively-designed RET based on the ViT architecture. The RET module enhances the fidelity of generated gestures through iterative refinement during inference. It employs channel-wise operations where hand and body features are processed separately in distinct channels, analogous to image processing frameworks. This disentanglement of body parts enables precise anatomical differentiation while endowing RET with more comprehensive spatial information processing capabilities. For detailed information, please refer to \cref{sec:GestureGenerator}.

The second improvement is the implementation of the key data processing strategy. It is Dynamic Embedding Regularization (DER). This method applies dropout\cite{dropout,dropoutNet} operations after the embedding layer. The dropout mechanism remains active during training but is disabled during inference. This is crucial for enhancing the model's learning ability, as it reduces the complex co-adaptations \cite{dropout} in large models and meanwhile introduces more noise to enhance the model's robustness. Moreover, thanks to the characteristic of this strategy that helps in reducing overfitting, it effectively improves the generalization ability of the model. In addition, in order to adapt to the IRI strategy and enhance the model's robustness, we applied the ``masking" operation to the ground truth (GT). Specifically, we randomly mask some GT data and then train the RET model to predict the clean GT, as shown in \cref{fig:EDIT}. 

To further overcome the disadvantages of autoregressive inference like error accumulation and the lack of self-correcting mechanisms, we propose the third improvement to optimize the inference phase. During inference, we apply Iterative Reconstruction Inference (IRI) to ensure the effectiveness of the masking strategy. Unlike the training phase, the IRI strategy starts with completely masked indices (indices are the positions within a codebook). It predicts indices with high confidence, and iteratively uncovers the remaining masked indices. For a more detailed discussion on the IRI strategy, please refer to \cref{sec:Inference}. To further enhance the model performance and the visual perception effect of the generated results, we introduce Classifier-Free Guidance (CFG) \cite{ClassifierFree} and the temporal smoothing process. For more details, please refer to \cref{sec:TrainingDetail} and \cref{sec:SmoothingProcess}.

After implementing the aforementioned strategies, our model outperforms those of previous work from the experimental results in \cref{sec:Comparisons}. Our results demonstrate that our method exhibits better and more stable generation performance when applied in practice.

In summary, our contributions are as follows:

\begin{itemize}
	\item Our work leverages the structural characteristics of ViT to design the RET. By incorporating channel-wise operations, the model effectively perceives and processes spatio-temporal information. Additionally, we preserve the inherent advantages of the ViT model, specifically its scalability and compatibility with other models. Our work provides certain theoretical support for the selection of model structures in the field of gesture generation.
	
	\item By deploying the DER strategy, we significantly enhance the model's learning capability. Moreover, thanks to the DER strategy's characteristic of reducing overfitting, the DER strategy effectively improves the generalization ability of the model. Thereby, the model avoids the generation of frozen movements or penetration in animation when handling out-of-domain audio inputs.
	
	\item Previous methods usually use autoregressive inference. Since it relies too heavily on previously generated data, such heavy reliance often leads to error accumulation and a lack of self-correcting mechanisms. To address these limitations, we propose a novel inference strategy (IRI), adopt the CFG strategy with prudence, and explore a temporal smoothing process suitable for Transformer-encoder models. These strategies have a positive impact on the model.
	
\end{itemize}
\section{Related Work}

In early co-speech gesture generation research, rule-based systems were commonly used. These methods defined manually-made rules to synchronize speech and gestures. Specific speech elements triggered corresponding gestures, like a pointing gesture for object-related nouns or a waving one for greetings, making gestures highly semantic. However, they had limitations. They demanded much manpower for rule design and maintenance, and lacked flexibility, struggling to adapt to diverse speaking styles and contexts. Relevant literature comprehensively reviews these methods \cite {RuleBaseReview1,RuleBaseReview2,RuleBaseReview3}.

To alleviate the problems existed in the rule-based methods mentioned above, a large number of data-driven methods are proposed. \cite{Hasegawa} uses a bidirectional Long Short-Term Memory to generate gestures from audio utterances. By learning the audio-gesture relationships, it predicts the full-skeleton human postures at each time step and then performs temporal filtering to smooth the posture sequences. \cite{Kucherenko} extends the work of \cite{Hasegawa}. It removes the need for temporal smoothing through representation learning of autoencoders and converts audio input into gesture sequences in the form of 3D joint coordinates. \cite{Ferstl1, Ferstl2} adds multiple adversarial objectives to recurrent neural networks to deal with the problem of regression to the mean that is likely to occur in long sequences. They also consider the phase structure of gestures, motion realism, displacement and the diversity of mini-batches. \cite{Ahuja} learns to implicitly estimate the mixture density to achieve the transfer of different speaker styles. Through adversarial training, the generated gestures can represent the styles of different speakers. \cite{Alexanderson} extends the normalization-flow-based model to speech-driven gesture synthesis, achieving style control. The generated gestures perform well in terms of naturalness and appropriateness. \cite{Habibie} presents a speech-driven 3D gesture synthesis approach. It has a generator and discriminator. The generator converts audio features to 3D pose sequences. Trained with regression and adversarial losses, it generates gestures in sync with speech. \cite{audio2gesture} uses conditional VAE to convert speech into diverse gestures. It models the speech-gesture mapping by splitting the cross-modal latent code into shared code and motion-specific code. \cite{Taylor} adopts the conditional Flow-VAE framework to generate the spontaneous gestures of the speaker and listener roles in conversations. It trains the model through an autoregressive framework and improves the expressiveness of the generated gestures. \cite{Zhou} generates gesture videos that match unseen audio inputs by conducting audio-video searches in the video gesture database. It uses video motion graphs and the beam search algorithm to find the optimal order of gesture frames.

Recently, \cite{gesturediffuclip} uses multi-modal inputs and latent diffusion models. It learns motion via VQ-VAE and synthesizes gestures in the latent space. With contrastive learning and AdaIN layer, it realizes flexible style-controlled co-speech gesture synthesis. \cite{emage} presents EMAGE to generate full-body gestures from audio and masked gestures. It creates BEAT2 and applies specific techniques for joint optimization, promoting the field.
\section{Pipeline Overview}
\label{sec:pipelineOverview}
\begin{figure}
	\centerline{\includegraphics[width=\columnwidth]{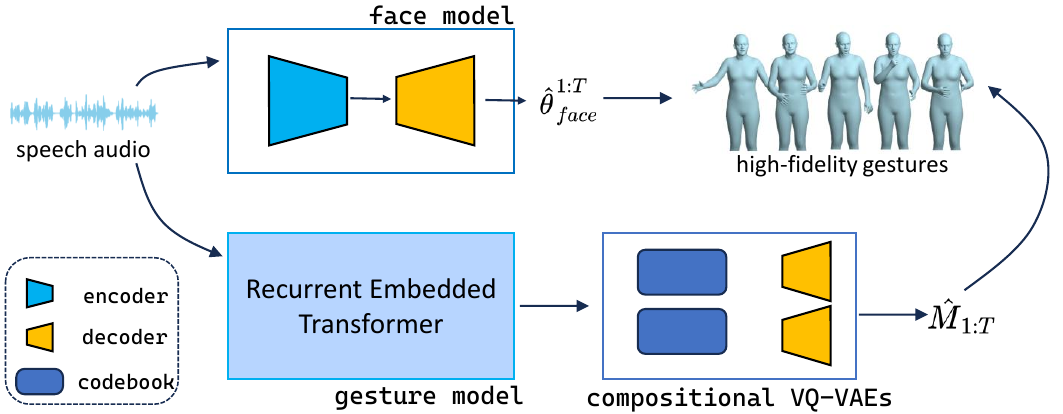}}
	\caption{Our ReCoM pipeline consists of three components: a face generator, a gesture generator and a compositional VQ-VAE. By inputting speech audio, we can obtain the corresponding gesture sequence. Among them, the face model is responsible for generating the facial movement sequence $\hat{\theta}_{face}^{1:T}$. The gesture generator aims to generate gesture indices with high confidence. These indices are then input to the compositional VQ-VAE for nearest neighbor search and decoding to obtain the gesture sequence $\hat{M}_{1:T}$.}
	\label{fig:systemoverview}
\end{figure}

Given a speech recording, the goal of our ReCoM is to generate high-fidelity gestures corresponding to it. The overall pipeline is depicted in \cref {fig:systemoverview}. In our method, we use $\theta_{face}\in \mathbb{R}^{103}$ to represent the whole face parameter. It consists of $\theta_{j}$ and $\theta_{e}$, where $\theta_{j} \in \mathbb{R}^{3} $ represents jaw pose and $\theta_{e} \in \mathbb{R}^{100}$ represents FLAME \cite{FLAME} expression parameters. $M_{1:T}$ denotes a $T$ frame gesture clip, and $\hat{M}_{1:T}$ represents the corresponding reconstructed gesture clip. $T$ is the number of the fixed frames used during training. $M^{b}$ and $M^{h}$ represent body pose and hand pose. $E_{1:t} = \{e_{1} , . . . , e_{t} \} \in \mathbb{R}^{t \times 64}$ denotes codebook vectors, and $Z_{1:t} = \{z_{1} , . . . , z_{t} \}  \in \mathbb{R}^{t \times 64}$ denotes latent vectors, and $t=\frac{T}{4}=22$. $A_{1:T}$ is the MFCC feature of audio. And $id$ represents the speaker's identity drawn from a predefined identity set.
\section{Method}

\subsection{Face Generator}\label{sec:FaceGenerator}

\begin{figure}
	\centerline{\includegraphics[width=\columnwidth]{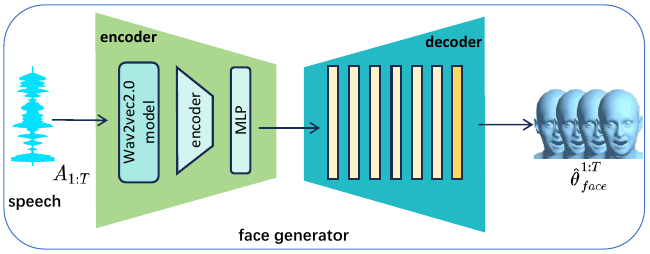}}
	\caption{For face generator, we choose encoder-decoder architecture.}
	\label{fig:FaceGenerator}
\end{figure}

For the face, following \cite{talkshow}, we adopt an encoder-decoder architecture, as illustrated in \cref{fig:FaceGenerator}. Owing to the excellent capability of learning audio representation, we adopt wav2vec 2.0 \cite{wavevec2} as the backbone. The decoder consists of multiple temporal convolution layers. We construct the face parameter distribution through loss function as follows:
\begin{equation}
	L_{face} = L_{jaw}(\theta_{j}, \hat{\theta}_{j}) + L_{expression}(\theta_{e}, \hat{\theta}_{e}),
\end{equation}
where $L_{jaw}$ and $L_{expression}$ are the L1 and L2 reconstruction losses. $\hat{\theta}$ denotes the corresponding reconstructed feature.

\subsection{Gesture Codebook}
\label{sec:GestureCodebook}

In the process of reconstructing and generating the hand and upper body parts, we use a \textbf{two-stage} approach. In the \textbf{first stage} we use VQ-VAE \cite{VQ-VAE}, which can learn a discrete representation and ensure that the poses are accurately reconstructed through latent vectors.

Our goal in this stage is to train VQ-VAE well enough to reconstruct pose data, preparing it for the generation task in the next stage. We use a loss function $L_{VQ}$ that includes reconstruction loss $L_{rec}$, codebook loss $L_{eq}$, and reconstruction speed loss $L_{v}$ as follows:
\begin{equation}\resizebox{0.89\hsize}{!}{$	
	L_{VQ} = L_{rec}(M_{1:T},\hat{M}_{1:T}) + L_{eq}(Z_{1:t},E_{1:t}) + L_{v}(M_{1:T},\hat{M}_{1:T})$},
\end{equation}

\begin{equation}	
	L_{eq} =  \|sg[E_{1:t}] - Z_{1:t} \| + 0.25 \|E_{1:t} - sg[Z_{1:t}]\|,
\end{equation}

\begin{equation}\resizebox{0.89\hsize}{!}{$	
	L_{v} = \frac{1}{T-1} \sum_{i}^{T-1} \mid(M_{1:T-1}-M_{2:T}) - (\hat{M}_{1:T-1} -\hat{M}_{2:T})\mid $},
\end{equation}
where \textbf{sg} denotes stop gradient operation. $L_{rec}$ is Mean Absolute Error loss. We learn two VQ-VAE for the hand and upper body, with their codebooks respectively designated as $C_{h}$ and $C_{b}$. The $C_{h}$ quantizes $M^{h} \in \mathbb{R}^{T \times 90}$, and the $C_{b}$ quantizes $M^{b} \in \mathbb{R}^{T \times 63}$. In training phase, we first encode the inputs $M_{1:T}$ into an embedding $Z_{1:t}$, then we use a Nearest Neighbor Search to return codebook indices. Lastly, we retrieve the corresponding embedding $E_{1:t}$ for input to the decoder to reconstruct the poses $\hat{M}_{1:T}$. We train the model using the loss function $L_{VQ}$.

\subsection{Gesture Generator}
\label{sec:GestureGenerator}

\begin{figure}
	\centerline{\includegraphics[width=\columnwidth]{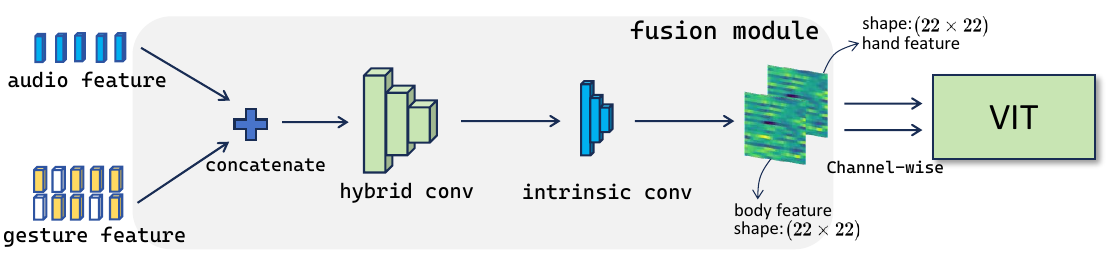}}
	\caption{Fusion module. We apply hybrid convolution (with a downsampling rate of 2 and a convolution kernel of $1 \times 1$) to fuse audio features and gesture features. The role of hybrid convolution here is to combine the audio and gesture features, enabling them to interact and form a unified feature representation. Then, we use intrinsic convolution to obtain the essential mixed features. Intrinsic convolution serves to downsample the mixed features into the latent space. This downsampling operation is crucial as it allows the ViT module to process the data without the need for an overly large number of parameters. Finally, we input the features into the ViT model by adopting a channel-wise strategy.}
	\label{fig:fusion}
\end{figure}

The \textbf{second stage} focuses on generating high-fidelity gestures and enhancing the generalization of the model. The gesture generator's core is the ViT \cite {ViT}. To save training time, we train the generator in the indice space of the codebook, with the loss function employing only \textbf{cross-entropy loss} as follows:

\begin{equation}	\resizebox{0.89\hsize}{!}{$
	L_{cls} = CrossEntropy(I_{1:t}, ViT(fusion(\overset{m}{I_{1:t}} ,a_{1:t} ),id))
	\label{eqn:GeneratorLoss} $},
\end{equation}
where $I \in \mathbb{R}^{t \times 2}$ denotes the indices of poses. $fusion$ module is shown in \cref{fig:fusion}. $a_{1:t}$ represents the audio feature after downsampling. $\overset{m}{I_{1:t}}$ denotes the indices of poses being masked.

Due to the challenge of establishing a mapping from audio to gestures, depending solely on audio input complicates the convergence of training. Therefore, to accelerate loss convergence, we incorporate the GT pose as an additional input for auxiliary. Thus, our gesture generator is fed with $A_{1:T}$ and GT pose $M_{1:T}$ data in the training phase.

As shown in \cref{fig:EDIT}, we first use two codebook encoders to get pose indices $I_{1:t}$ and use an audio encoder and $1 \times 1$ convolution to get audio feature $a_{1:t}$. Then, we apply a masking strategy (like \cite{bert}) to $I_{1:t}$ and pass the masked $I_{1:t}$ through an embedding layer to map $I_{1:t}$ and $a_{1:t}$ to the same dimensions. For pose features, we employ a dropout strategy to enhance the model's learning capability by introducing random perturbations. Later, we fuse the audio and pose features using a fusion module and then input them into the ViT model. Notably, the input operation is carried out channel-wise, splitting the body and hand poses into two channels (injecting spatial information into the third dimension of the features). This is similar to an RGB image (shape is width $\times$ height $\times$ 2), where each channel represents different components yet is closely interconnected. As shown in \cref{fig:fusion}, we divide the hand features and body features into two feature maps. Meanwhile, the width and height of the feature maps correspond to the temporal and spatial dimensions of the gesture clips respectively. For the spatial dimension corresponding to the width, we do not perform compression, while for the feature dimension corresponding to the height, we conduct downsampling using intrinsic convolution. Thus, channel-wise processing is crucial for the effectiveness of the RET model, as it is essential to fuse spatio-temporal information.

In the ViT model, we first apply patchEmbed to different channels of the input data. Then, we add $id$ and position encoding, both of which are fed into $N$ ViT blocks ($N$ is 15) to obtain the predicted indices $\hat{I}$. Finally, by applying the loss function to $I$ and $\hat{I}$, we can train the model to converge.

\subsection{Training Detail}
\label{sec:TrainingDetail}

In our experiments, we find that if we only input speech data, the model is hard to converge. Thus, we combine speech and GT pose as input data. But, if the GT pose isn't processed, the model is prone to overfitting. After extensive experimentation, we find that DER and masking strategies can help to alleviate overfitting. These strategies introduce large random perturbations to the input data, not only guiding the model to learn more robust features, but also improving the generalization ability of the model. Additionally, we employ channel-wise operation to fuse spatio-temporal information, and apply an Exponential Moving Average (EMA) technique during model training. 

We use CFG \cite{ClassifierFree} to train our model. Firstly, we try to disable the speech $a_{1:t}$ (i.e.\ Empty condition) in \cref{eqn:GeneratorLoss} with a probability of 10\% like \cite{ClassifierFree}, but it fails to work as expected. Thus, we directly employ the dropout operation to substitute the Empty condition, which effectively enhances the performance of the model. During inference, we use the following \cref{eqn:guider} in the last neural network layer before softmax to guide the generation process:

\begin{equation}	
	logit = s \cdot RET(a_{1:t} , id) - (s-1) \cdot RET(\phi ,id),
	\label{eqn:guider}
\end{equation}
where $s$ is guidance scale, and $logit$ is the guided generation result. We can control the speaker gesture style through $id$. It is worth noting that, in order to achieve good results, the linear combination in \cref{eqn:guider} we adopted is different from that in \cite{ClassifierFree}. In addition, we conduct experiments, as shown in \cref{tab:VerifyingCFG}, to prove that in our model, it is better to use dropout than the Empty condition during training. As the results show, among these three methods (A, B, and C), when the CFG inference is utilized, employing the Empty condition or not processing the speech condition has a negative effect on the model performance. Using the dropout operation in combination with the CFG inference instead has a positive effect.

\begin{table}
	\caption{\textbf{A} represents training with the Empty condition at a certain probability. \textbf{B} denotes training with the dropout operation at a certain probability. \textbf{C} represents using neither of the two ways, i.e., not conducting any processing on the speech condition. \textbf{EN} represents enabling \cref{eqn:guider} during inference, while \textbf{UN} represents not enabling.}
	\label{tab:VerifyingCFG}
	\begin{minipage}{\columnwidth}
		\begin{center}
			\resizebox{\textwidth}{!}{
			\begin{tabular}{ r c c c c }
				\toprule
				\textbf{Method} & \textbf{\textit{Diversity$\uparrow$}}& \textbf{ \textit{FGD$\downarrow$}} & \textbf{\textit{MAE$\downarrow$}} & \textbf{\textit{BC$\rightarrow$}} \\ \midrule
				A and UN&8.4988&30.1020 &35.5114&0.8570 \\
				A and EN&9.3009&100.1004 &35.8544&0.8569 \\
                \midrule
				B and UN&8.2614&10.8462 &\textbf{35.4285}&0.8574 \\
				B and EN&8.9830&\textbf{2.4816}&35.9665&\textbf{0.8579} \\
                \midrule
				C and UN&8.3830&16.7449 &35.4646&0.8567 \\
				C and EN&\textbf{10.9710}&143.0840&36.6753&0.8578 \\
				\bottomrule
			\end{tabular}
		}
		\end{center}
	\end{minipage}
\end{table}
\section{Inference Phase}\label{sec:Inference}

In the inference phase, we divide the process into two parts: face inference and gesture inference. For face inference, we have the VAE architecture as mentioned in \cref{sec:FaceGenerator}. Given an audio input, the facial model trained on facial distributions can generate a FLAME \cite{FLAME} result.

\subsection{Iterative Reconstruction Inference}
For the gesture inference part, we propose a novel inference strategy, termed IRI, to enhance the generation results, as illustrated in \cref{fig:EDIT}. Specifically, the method initially takes speech features and fully masked motion indices as inputs. Then, we repeatedly input the indices and speech features into the RET model to predict masked indices until all the indices are recovered. In this process, the model automatically selects results that exceed the confidence threshold value. Indices with results below the threshold will be predicted in next iterations. The threshold value decreasing adaptively in a linear manner aims at reducing the difficulty of data reconstruction. This adjustment is necessary because some of the data, due to the accumulation of global errors, cannot achieve results with high confidence. Making predictions completely out of chronological order helps alleviate the cumulative errors in the time sequence. Moreover, by re-inputting the low-confidence indices into the RET model for prediction, the model can correct the low-confidence results it generates.

\subsection{Temporal Smoothing Process}
\label{sec:SmoothingProcess}
Transformer encoder models \cite{Transformer} require the length of sequences to be strictly equal to the fixed length during training. To generate long gesture sequences, we need to concatenate different result segments. Initially, we use a direct concatenation strategy. However, this results in incoherence at the junctions of the generated outputs, similar to \cite{ProbTalk}. To address this issue, we need to suppress the excessive freedom of the RET model, guide the generation direction, and produce results that align more closely with human visual perception. With these goals in mind, we propose the Smoothing strategy as shown in \cref{fig:EDIT}. Surprisingly, this temporal smoothing process significantly improves the visual coherence of the overall results, not just at the connection points of the results. Specifically, we divide the speech audio into several short segments in chronological order. For an audio with frame rate of 30FPS, we split it into segments with 88 frames each. For the segments $S_{i}$ other than the first one, we incorporate the last 8 frames of segment $S_{i-1}$ into the segment $S_{i}$. This enables the information of several independent segments to be transmitted in a temporal sequence.

\subsection{Editing Capabilities}
Since we adopt the Transformer-encoder as the main architecture of the gesture generator, our RET model acquires certain editing capabilities. This means that by adding a target pose, it can generate gestures that match the target gesture. In addition, our model can smoothly splice two gesture clips together. Moreover, similar to the approach in \cite{coma} which uses residual VQ-VAE \cite{RQVAE}, we are able to independently control body or hand poses based on the characteristics of the VQ-VAE \cite{VQ-VAE}. Please refer to supplementary material for editing examples.
\section{Experiment}

\subsection{Quantitative Comparisons}\label{sec:Comparisons}

\begin{figure*}
	\centerline{\includegraphics[width=\textwidth]{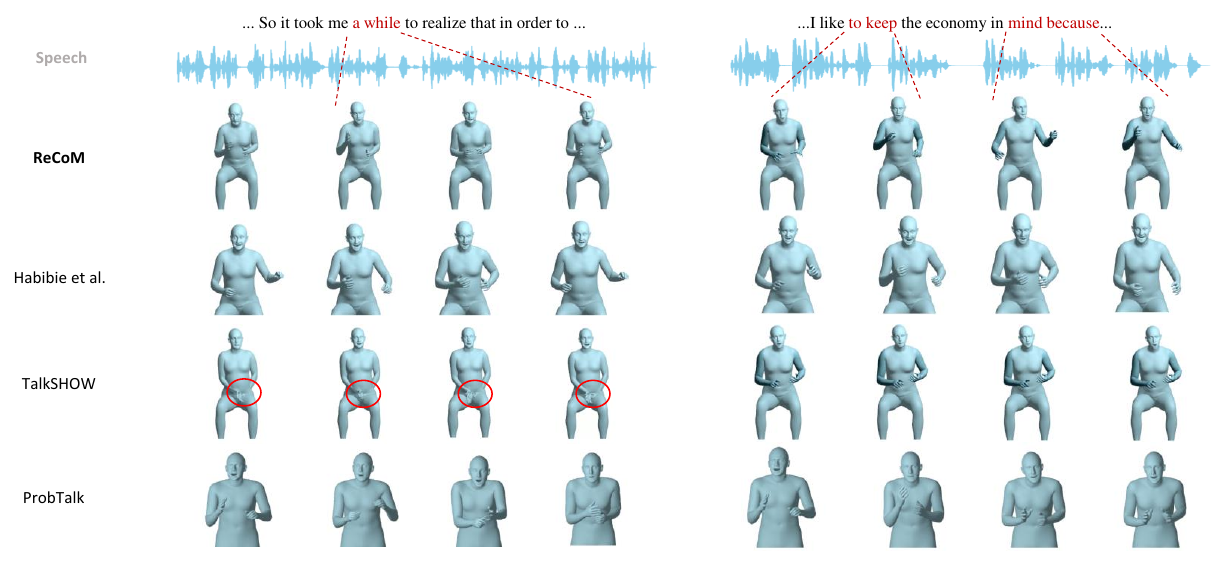}}
	\caption{When receiving out-of-domain audio inputs, TalkSHOW exhibits a frozen motion for a number of seconds. The results of ProbTalk often show incoherence between two consecutive frames. Meanwhile, Habibie et al.'s method often generates gestures with overly large jitter amplitudes. Our ReCoM results instead remain natural. Please refer to supplementary materials for more video demos, with both in-domain and out-of-domain evaluation.}
	\label{fig:Qualitative}
\end{figure*}

We compare ReCoM with \cite{Habibie}, TalkSHOW \cite{talkshow} and ProbTalk \cite{ProbTalk}. We select a series of metrics to ensure a comprehensive and accurate assessment of the model's performance across multiple aspects. 

\begin{table}
	\caption{In-domain evaluation on the \textbf{SHOW} dataset. The downward arrow indicates that smaller values are better, and vice versa for the upward arrow. The right-pointing arrow indicates that the closer the value is to the ground truth, the better. Bold and underlined indicate the best and the 2nd best results. And GT denotes Ground Truth.}
	\label{tab:QuantitativeTable}
	\begin{minipage}{\columnwidth}
		\begin{center}
			\resizebox{\textwidth}{!}{
			\begin{tabular}{ r c c c c }
				\toprule
				\textbf{Method} & \textbf{\textit{Diversity$\uparrow$}}& \textbf{ \textit{FGD$\downarrow$}} & \textbf{\textit{MAE$\downarrow$}} & \textbf{\textit{BC$\rightarrow$}} \\ \midrule
				
				GT&9.4850&0 &0&0.8676 \\
				\hline
				
				Habibie et al.&7.5246&239.1780 &98.6942&0.9477 \\
				
				TalkSHOW&6.8678&66.1574&36.7540& \textbf{0.8713} \\
				
				ProbTalk&\underline{7.6758}&\underline{18.7028}&\underline{36.0005}& 0.7837 \\
				
				ReCoM& \textbf{8.9830}&\textbf{2.4816} &\textbf{35.9665}&\underline{0.8579} \\
				\bottomrule
			\end{tabular}
		}
		\end{center}
	\end{minipage}
\end{table}

\begin{table}
	\caption{Out-of-domain evaluation on \textbf{BEAT2} without any fine-tuning.}
	\label{tab:generalization}
	\begin{minipage}{\columnwidth}
		\begin{center}
			\resizebox{\textwidth}{!}{
			\begin{tabular}{ r c c c c }
				\toprule
				\textbf{Method} & \textbf{\textit{Diversity$\uparrow$}}& \textbf{ \textit{FGD$\downarrow$}} & \textbf{\textit{MAE$\downarrow$}} & \textbf{\textit{BC$\rightarrow$}} \\ \midrule
				GT&14.8500&0 &0&0.8351 \\
				\hline
				Habibie et al.&7.5242&239.1844 &92.2333&0.9477 \\
				TalkSHOW&\underline{8.6990}&\underline{98.3199} &72.2534&0.8729 \\
				ProbTalk&8.2616&100.0674&\underline{71.6509}&\underline{0.8178} \\
				ReCoM& \textbf{11.1303}&\textbf{96.7793} &\textbf{71.5830}&\textbf{0.8469} \\
				\bottomrule
			\end{tabular}
		}
		\end{center}
	\end{minipage}
\end{table}

\textbf{Diversity.} The metric follows \cite{emage} and represents the average L1 distance between $K$ gesture clips. Its calculation formula is shown in \cref{eq:diversityEq}: 

\begin{equation}
	L_1^{div} = \frac{1}{2K(K-1)} \sum_{t=1}^{K} \sum_{i}  \|p_{t}^{i}- \bar{p}_{t} \|_{1},
    \label{eq:diversityEq}
\end{equation}
where $p_{t}^{i}$ denotes the position of joints in frame $i$ of the $t$-th gesture clip. Where $\bar{p}_{t}$ represents the average value of the joint positions of the $t$-th gesture clip.

\textbf{FGD (Frechet Gesture Distance).} Similar to the FID metric, it is an important metric for measuring the similarity between two multi-dimensional Gaussian distributions, particularly suitable for assessing the performance of generative models at the feature space \cite{FGD,EvaluatingFGD}. A smaller FGD value indicates that the distribution of generated samples is closely aligned with the distribution of real samples, suggesting high fidelity of the model's outputs.

\textbf{MAE (Mean Absolute Error).} It calculates the average of the sum of absolute differences between real and generated samples in the feature space.

\textbf{BC (Beat Consistency Score).} The objective of the BC metric is to gauge the correlation between the variation degree of human joint points and audio beats \cite{BC}.

Experimental results, as reported in \cref{tab:QuantitativeTable} and \cref{tab:generalization}, demonstrate our model's high fidelity and good generalization. The higher diversity metrics are attributed to non-autoregressive inference. That is, the model doesn't rely on previously generated data, which improves the diversity. In addition, lower FGD and MAE metrics indicate that our model better fits the training dataset. The lower Beat Consistency (BC) metric might be attributed to the semantic association ability of our model. For instance, it can link the sound of clapping with the actual clapping motion, associate the word "no" with the action of quickly opening the arms, and connect "hip hop" to similar dance styles. Precisely because of this semantic association ability, the model tends to focus less on its own beat consistency, which in turn results in a lower BC metric.

\subsection{Qualitative Comparisons}

As shown in \cref{fig:Qualitative}, we present two gesture clips generated from an out-of-domain audio segment, displaying frame captures that span approximately four seconds. It can be observed that our method is capable of performing inference as usual when receiving out-of-domain audio inputs. However, due to the autoregressive method used in TalkSHOW \cite{talkshow}, this results in error accumulation and a lack of self-correcting mechanisms, making it prone to producing frozen motion and occasional penetration in animation. Additionally, TalkSHOW often outputs the unnatural motions, e.g., keeping avatar's left hand to one side. Due to our model's non-autoregressive architecture, we don't have these problems.

\begin{figure}
	\centerline{\includegraphics[width=\columnwidth]{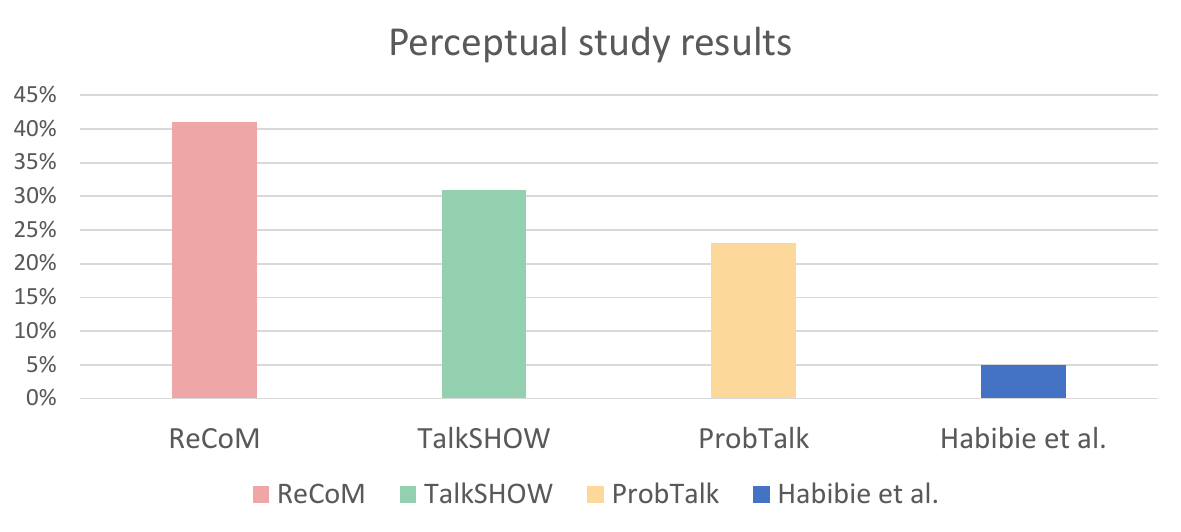}}
	\caption{Perceptual study results. We calculate the win rate of the evaluation. Our method demonstrate higher overall preference of participants.}
	\label{fig:PerceptualStudy}
\end{figure}

\textbf{Perceptual Study.} Objective metrics do not always reflect the model's performance. Therefore, to further verify the visual performance of our ReCoM method, we conducted a perceptual study. In total, we sampled 81 videos. Among them, 50 were from the SHOW test set, while the remaining 31, including 5 long ones with a length of 60 seconds, were from the wild TED audio. Twenty participants evaluated videos generated by different methods presented in random order, selecting their most preferred option. The results are shown in the \cref{fig:PerceptualStudy}, and our method demonstrates higher overall preference among participants. Notably, although ProbTalk has better objective metrics than TalkSHOW, it achieved somewhat worse results in the perceptual study. We attribute this to the fact that the motions generated by ProbTalk are not smooth, often resulting in the phenomenon of "frame-skipping", where there is a lack of coherence between consecutive frames in the video. We believe this may have a significant impact on the visual experience of the subjects.

\subsection{Ablation Study}\label{Ablation}

\begin{table}
	\caption{Our ablation results, which is tested on the SHOW dataset. \textbf{w/o} denotes ``without''. \textbf{DER} means ``Dynamic Embedding Regularization''.}
	\label{tab:AblationTable}
	\begin{minipage}{\columnwidth}
		\begin{center}
			\resizebox{\textwidth}{!}{
			\begin{tabular}{ r c c c c }
				\toprule
				& \textbf{\textit{Diversity$\uparrow$}}& \textbf{ \textit{FGD$\downarrow$}} & \textbf{\textit{MAE$\downarrow$}} & \textbf{\textit{BC$\rightarrow$}} \\ \midrule
				ReCoM&\textbf{8.9830}&\textbf{2.4816}&35.9665&\textbf{0.8579} \\
				w/o CFG&8.2614&10.8462&35.4285&0.8574 \\
				w/o IRI&8.7314&39.9367 &\textbf{31.7857}&0.8570 \\
				w/o audio-dropout&8.3830&16.7449&35.4646&0.8567 \\
				w/o EMA  &8.1029 & 27.6172 & 35.4365 & 0.8570 \\
				w/o DER&6.9025&146.3948 &35.2953&0.8545 \\
				w/o masking&8.4321&71.0111 &35.6858&0.8560 \\
				\bottomrule
			\end{tabular}
		}
		\end{center}
	\end{minipage}
\end{table}

As shown in \cref{tab:AblationTable}, to validate the effectiveness of our design, we conduct ablation experiments by removing these strategies. It can be observed that these strategies all have a significant impact on the FGD metric. Although the RET model isn't the best in all metrics, we choose it because FGD has the biggest impact on visual effect.
\section{Conclusion, Limitation and Future Work}

We design the RET by retaining the structural characteristics of ViT, and enable the model to process spatio-temporal information through channel-wise operations. Additionally, we apply some strategies to improve the model's convergence procedure, and we use new inference strategy to enhance the generation results. This method achieves significant results, not only enhancing the model's performance but also improving its generalization ability. Finally, our model's design enables it to have certain editing capabilities.

It must be acknowledged that our model has a deficiency in terms of beat consistency, but we believe this trade-off is worthwhile. Furthermore, our work does not take into account lower body motion, resulting in a lack of overall body information. This can cause occasional penetration between the hand and lower body in seated poses, reducing the animation's practicality. However, model penetration will hardly occur in the upper body and hands. Lastly, due to the self-correcting mechanism in our model, it occasionally generates faster movements as it attempts to correct errors made earlier.

In future work, we may try to combine multi-modal information, such as text, to enhance the semantic results of our model. Additionally, we hope to design a model capable of generating whole body gestures of high fidelity, including the lower body, which will be challenging yet highly practical. Meanwhile, we observe that the existing metrics can hardly fully test the performance of the model from a human perspective. Thus, we will attempt to propose more comprehensive evaluation metrics.
{
    \small
    \bibliographystyle{ieeenat_fullname}
    \bibliography{main}

\begin{thebibliography}{53}
\providecommand{\natexlab}[1]{#1}
\providecommand{\url}[1]{\texttt{#1}}
\expandafter\ifx\csname urlstyle\endcsname\relax
  \providecommand{\doi}[1]{doi: #1}\else
  \providecommand{\doi}{doi: \begingroup \urlstyle{rm}\Url}\fi

\bibitem[Ahuja et~al.(2020)Ahuja, Lee, Nakano, and Morency]{Ahuja}
Chaitanya Ahuja, Dong~Won Lee, Yukiko~I. Nakano, and Louis-Philippe Morency.
\newblock Style transfer for co-speech gesture animation: A multi-speaker
  conditional-mixture approach.
\newblock In \emph{European Conference on Computer Vision}, 2020.

\bibitem[Alexanderson et~al.(2020)Alexanderson, Henter, Kucherenko, and
  Beskow]{Alexanderson}
Simon Alexanderson, Gustav~Eje Henter, Taras Kucherenko, and Jonas Beskow.
\newblock Style‐controllable speech‐driven gesture synthesis using
  normalising flows.
\newblock \emph{Computer Graphics Forum}, 39, 2020.

\bibitem[Alexanderson et~al.(2023)Alexanderson, Nagy, Beskow, and
  Henter]{ListenAction}
Simon Alexanderson, Rajmund Nagy, Jonas Beskow, and Gustav~Eje Henter.
\newblock Listen, denoise, action! audio-driven motion synthesis with diffusion
  models.
\newblock \emph{ACM Trans. Graph.}, 42\penalty0 (4), 2023.

\bibitem[Ao et~al.(2023)Ao, Zhang, and Liu]{gesturediffuclip}
Tenglong Ao, Zeyi Zhang, and Libin Liu.
\newblock Gesturediffuclip: Gesture diffusion model with clip latents.
\newblock \emph{ACM Transactions on Graphics (TOG)}, 42\penalty0 (4):\penalty0
  1--18, 2023.

\bibitem[Baevski et~al.(2020)Baevski, Zhou, Mohamed, and Auli]{wavevec2}
Alexei Baevski, Henry Zhou, Abdelrahman Mohamed, and Michael Auli.
\newblock wav2vec 2.0: A framework for self-supervised learning of speech
  representations.
\newblock \emph{arXiv: Computation and Language,arXiv: Computation and
  Language}, 2020.

\bibitem[Chen et~al.(2024)Chen, Liu, Wang, Zeng, Li, and Chen]{diffsheg}
Junming Chen, Yunfei Liu, Jianan Wang, Ailing Zeng, Yu Li, and Qifeng Chen.
\newblock Diffsheg: A diffusion-based approach for real-time speech-driven
  holistic 3d expression and gesture generation.
\newblock In \emph{Proceedings of the IEEE/CVF Conference on Computer Vision
  and Pattern Recognition}, pages 7352--7361, 2024.

\bibitem[Chhatre et~al.(2024)Chhatre, Dan??ek, Athanasiou, Becherini, Peters,
  Black, and Bolkart]{amuse}
Kiran Chhatre, Radek Dan??ek, Nikos Athanasiou, Giorgio Becherini, Christopher
  Peters, Michael~J. Black, and Timo Bolkart.
\newblock Emotional speech-driven 3d body animation via disentangled latent
  diffusion.
\newblock In \emph{Proceedings of the IEEE/CVF Conference on Computer Vision
  and Pattern Recognition (CVPR)}, pages 1942--1953, 2024.

\bibitem[Devlin et~al.(2019)Devlin, Chang, Lee, and Toutanova]{bert}
Jacob Devlin, Ming-Wei Chang, Kenton Lee, and Kristina Toutanova.
\newblock {BERT}: Pre-training of deep bidirectional transformers for language
  understanding.
\newblock In \emph{Proceedings of the 2019 Conference of the North {A}merican
  Chapter of the Association for Computational Linguistics: Human Language
  Technologies, Volume 1 (Long and Short Papers)}, pages 4171--4186,
  Minneapolis, Minnesota, 2019. Association for Computational Linguistics.

\bibitem[Ferstl et~al.(2019)Ferstl, Neff, and Mcdonnell]{Ferstl1}
Ylva Ferstl, Michael Neff, and Rachel Mcdonnell.
\newblock Multi-objective adversarial gesture generation.
\newblock \emph{Proceedings of the 12th ACM SIGGRAPH Conference on Motion,
  Interaction and Games}, 2019.

\bibitem[Ferstl et~al.(2020)Ferstl, Neff, and Mcdonnell]{Ferstl2}
Ylva Ferstl, Michael Neff, and Rachel Mcdonnell.
\newblock Adversarial gesture generation with realistic gesture phasing.
\newblock \emph{Comput. Graph.}, 89:\penalty0 117--130, 2020.

\bibitem[Gal and Ghahramani(2015)]{dropoutNet}
Yarin Gal and Zoubin Ghahramani.
\newblock A theoretically grounded application of dropout in recurrent neural
  networks.
\newblock In \emph{Neural Information Processing Systems}, 2015.

\bibitem[Goodfellow et~al.(2020)Goodfellow, Pouget-Abadie, Mirza, Xu,
  Warde-Farley, Ozair, Courville, and Bengio]{GANs}
Ian Goodfellow, Jean Pouget-Abadie, Mehdi Mirza, Bing Xu, David Warde-Farley,
  Sherjil Ozair, Aaron Courville, and Yoshua Bengio.
\newblock Generative adversarial networks.
\newblock \emph{Communications of the ACM}, 63\penalty0 (11):\penalty0
  139--144, 2020.

\bibitem[Guo et~al.(2024)Guo, Mu, Javed, Wang, and Cheng]{momask}
Chuan Guo, Yuxuan Mu, Muhammad~Gohar Javed, Sen Wang, and Li Cheng.
\newblock Momask: Generative masked modeling of 3d human motions.
\newblock In \emph{Proceedings of the IEEE/CVF Conference on Computer Vision
  and Pattern Recognition}, pages 1900--1910, 2024.

\bibitem[Habibie et~al.(2021)Habibie, Xu, Mehta, Liu, Seidel, Pons-Moll,
  Elgharib, and Theobalt]{Habibie}
Ikhsanul Habibie, Weipeng Xu, Dushyant Mehta, Lingjie Liu, Hans-Peter Seidel,
  Gerard Pons-Moll, Mohamed Elgharib, and Christian Theobalt.
\newblock Learning speech-driven 3d conversational gestures from video.
\newblock In \emph{Proceedings of the 21st ACM International Conference on
  Intelligent Virtual Agents}, pages 101--108, 2021.

\bibitem[Hasegawa et~al.(2018)Hasegawa, KANEKO, Shirakawa, Sakuta, and
  Sumi]{Hasegawa}
Dai Hasegawa, Naoshi KANEKO, Shinichi Shirakawa, Hiroshi Sakuta, and Kazuhiko
  Sumi.
\newblock Evaluation of speech-to-gesture generation using bi-directional lstm
  network.
\newblock 2018.

\bibitem[Ho and Salimans(2022)]{ClassifierFree}
Jonathan Ho and Tim Salimans.
\newblock Classifier-free diffusion guidance.
\newblock \emph{arXiv preprint arXiv:2207.12598}, 2022.

\bibitem[Ho et~al.(2020)Ho, Jain, and Abbeel]{DDPM}
Jonathan Ho, Ajay Jain, and Pieter Abbeel.
\newblock Denoising diffusion probabilistic models.
\newblock \emph{Advances in neural information processing systems},
  33:\penalty0 6840--6851, 2020.

\bibitem[Kingma(2013)]{VAE}
Diederik~P Kingma.
\newblock Auto-encoding variational bayes.
\newblock \emph{arXiv preprint arXiv:1312.6114}, 2013.

\bibitem[Kipp(2005)]{RuleBaseReview1}
Michael Kipp.
\newblock Gesture generation by imitation: from human behavior to computer
  character animation.
\newblock 2005.

\bibitem[Kolesnikov et~al.(2021)Kolesnikov, Dosovitskiy, Weissenborn, Heigold,
  Uszkoreit, Beyer, Minderer, Dehghani, Houlsby, Gelly, Unterthiner, and
  Zhai]{ViT}
Alexander Kolesnikov, Alexey Dosovitskiy, Dirk Weissenborn, Georg Heigold,
  Jakob Uszkoreit, Lucas Beyer, Matthias Minderer, Mostafa Dehghani, Neil
  Houlsby, Sylvain Gelly, Thomas Unterthiner, and Xiaohua Zhai.
\newblock An image is worth 16x16 words: Transformers for image recognition at
  scale.
\newblock 2021.

\bibitem[Kopp et~al.(2006)Kopp, Krenn, Marsella, Marshall, Pelachaud, Pirker,
  Th{\'o}risson, and Vilhj{\'a}lmsson]{RuleBaseReview2}
Stefan Kopp, Brigitte Krenn, Stacy Marsella, Andrew~N. Marshall, Catherine
  Pelachaud, Hannes Pirker, Kristinn~R. Th{\'o}risson, and Hannes~H{\"o}gni
  Vilhj{\'a}lmsson.
\newblock Towards a common framework for multimodal generation: The behavior
  markup language.
\newblock In \emph{International Conference on Intelligent Virtual Agents},
  2006.

\bibitem[Kucherenko et~al.(2019)Kucherenko, Hasegawa, Henter, Kaneko, and
  Kjellström]{Kucherenko}
Taras Kucherenko, Dai Hasegawa, Gustav Henter, Naoshi Kaneko, and Hedvig
  Kjellström.
\newblock Analyzing input and output representations for speech-driven gesture
  generation.
\newblock pages 97--104, 2019.

\bibitem[Kucherenko et~al.(2023)Kucherenko, Wolfert, Yoon, Viegas, Nikolov,
  Tsakov, and Henter]{EvaluatingFGD}
Taras Kucherenko, Pieter Wolfert, Youngwoo Yoon, Carla Viegas, Teodor Nikolov,
  Mihail Tsakov, and Gustav~Eje Henter.
\newblock Evaluating gesture generation in a large-scale open challenge: The
  genea challenge 2022.
\newblock \emph{ACM Transactions on Graphics}, 43:\penalty0 1 -- 28, 2023.

\bibitem[Lee et~al.(2022)Lee, Kim, Kim, Cho, and Han]{RQVAE}
Doyup Lee, Chiheon Kim, Saehoon Kim, Minsu Cho, and Wook-Shin Han.
\newblock Autoregressive image generation using residual quantization.
\newblock In \emph{Proceedings of the IEEE/CVF Conference on Computer Vision
  and Pattern Recognition (CVPR)}, pages 11523--11532, 2022.

\bibitem[Li et~al.(2021)Li, Kang, Pei, Zhe, Zhang, He, and Bao]{audio2gesture}
Jing Li, Di Kang, Wenjie Pei, Xuefei Zhe, Ying Zhang, Zhenyu He, and Linchao
  Bao.
\newblock Audio2gestures: Generating diverse gestures from speech audio with
  conditional variational autoencoders.
\newblock In \emph{2021 IEEE/CVF International Conference on Computer Vision
  (ICCV)}, pages 11273--11282, 2021.

\bibitem[Li et~al.(2017)Li, Bolkart, Black, Li, and Romero]{FLAME}
Tianye Li, Timo Bolkart, Michael~J Black, Hao Li, and Javier Romero.
\newblock Learning a model of facial shape and expression from 4d scans.
\newblock \emph{ACM Trans. Graph.}, 36\penalty0 (6):\penalty0 194--1, 2017.

\bibitem[Liang et~al.(2024)Liang, Bao, Zhang, Ren, Xu, Yang, Chen, Yu, and
  Xu]{omg}
Han Liang, Jiacheng Bao, Ruichi Zhang, Sihan Ren, Yuecheng Xu, Sibei Yang, Xin
  Chen, Jingyi Yu, and Lan Xu.
\newblock Omg: Towards open-vocabulary motion generation via mixture of
  controllers.
\newblock In \emph{Proceedings of the IEEE/CVF Conference on Computer Vision
  and Pattern Recognition}, pages 482--493, 2024.

\bibitem[Liu et~al.(2022{\natexlab{a}})Liu, Zhu, Iwamoto, Peng, Li, Zhou,
  Bozkurt, and Zheng]{beat}
Haiyang Liu, Zihao Zhu, Naoya Iwamoto, Yichen Peng, Zhengqing Li, You Zhou,
  Elif Bozkurt, and Bo Zheng.
\newblock Beat: A large-scale semantic and emotional multi-modal dataset for
  conversational gestures synthesis.
\newblock In \emph{European conference on computer vision}, pages 612--630.
  Springer, 2022{\natexlab{a}}.

\bibitem[Liu et~al.(2024{\natexlab{a}})Liu, Zhu, Becherini, Peng, Su, Zhou,
  Zhe, Iwamoto, Zheng, and Black]{emage}
Haiyang Liu, Zihao Zhu, Giorgio Becherini, Yichen Peng, Mingyang Su, You Zhou,
  Xuefei Zhe, Naoya Iwamoto, Bo Zheng, and Michael~J Black.
\newblock Emage: Towards unified holistic co-speech gesture generation via
  expressive masked audio gesture modeling.
\newblock In \emph{Proceedings of the IEEE/CVF Conference on Computer Vision
  and Pattern Recognition}, pages 1144--1154, 2024{\natexlab{a}}.

\bibitem[Liu et~al.(2022{\natexlab{b}})Liu, Wu, Zhou, Xu, Qian, Lin, Zhou, Wu,
  Dai, and Zhou]{BC}
Xian Liu, Qianyi Wu, Hang Zhou, Yinghao Xu, Rui Qian, Xinyi Lin, Xiaowei Zhou,
  Wayne Wu, Bo Dai, and Bolei Zhou.
\newblock Learning hierarchical cross-modal association for co-speech gesture
  generation.
\newblock In \emph{Proceedings of the IEEE/CVF Conference on Computer Vision
  and Pattern Recognition}, pages 10462--10472, 2022{\natexlab{b}}.

\bibitem[Liu et~al.(2024{\natexlab{b}})Liu, Cao, Wen, Jiang, and
  Ding]{ProbTalk}
Yifei Liu, Qiong Cao, Yandong Wen, Huaiguang Jiang, and Changxing Ding.
\newblock Towards variable and coordinated holistic co-speech motion
  generation.
\newblock In \emph{Proceedings of the IEEE/CVF Conference on Computer Vision
  and Pattern Recognition}, pages 1566--1576, 2024{\natexlab{b}}.

\bibitem[Peebles and Xie(2023)]{DiT}
William Peebles and Saining Xie.
\newblock Scalable diffusion models with transformers.
\newblock In \emph{Proceedings of the IEEE/CVF International Conference on
  Computer Vision (ICCV)}, pages 4195--4205, 2023.

\bibitem[Qi et~al.(2024)Qi, Zhang, Wang, Pan, Liu, Li, Chi, Li, Zhang, Xue,
  et~al.]{cocogesture}
Xingqun Qi, Hengyuan Zhang, Yatian Wang, Jiahao Pan, Chen Liu, Peng Li, Xiaowei
  Chi, Mengfei Li, Qixun Zhang, Wei Xue, et~al.
\newblock Cocogesture: Toward coherent co-speech 3d gesture generation in the
  wild.
\newblock \emph{arXiv preprint arXiv:2405.16874}, 2024.

\bibitem[Rombach et~al.(2021)Rombach, Blattmann, Lorenz, Esser, and Ommer]{LDM}
Robin Rombach, Andreas Blattmann, Dominik Lorenz, Patrick Esser, and
  Bj{\"{o}}rn Ommer.
\newblock High-resolution image synthesis with latent diffusion models.
\newblock \emph{CoRR}, abs/2112.10752, 2021.

\bibitem[Shi et~al.(2023)Shi, Luo, Peng, Zhang, and Sun]{fg-mdm}
Xu Shi, Chuanchen Luo, Junran Peng, Hongwen Zhang, and Yunlian Sun.
\newblock Fg-mdm: Towards zero-shot human motion generation via chatgpt-refined
  descriptions.
\newblock 2023.

\bibitem[Song et~al.(2021)Song, Meng, and Ermon]{DDIM}
Jiaming Song, Chenlin Meng, and Stefano Ermon.
\newblock Denoising diffusion implicit models.
\newblock In \emph{International Conference on Learning Representations}, 2021.

\bibitem[Srivastava et~al.(2014)Srivastava, Hinton, Krizhevsky, Sutskever, and
  Salakhutdinov]{dropout}
Nitish Srivastava, Geoffrey Hinton, Alex Krizhevsky, Ilya Sutskever, and Ruslan
  Salakhutdinov.
\newblock Dropout: a simple way to prevent neural networks from overfitting.
\newblock \emph{The journal of machine learning research}, 15\penalty0
  (1):\penalty0 1929--1958, 2014.

\bibitem[Sun et~al.(2025)Sun, Araujo, Xu, Zhou, Zhang, Huang, You, and
  Xie]{coma}
Shanlin Sun, Gabriel~De Araujo, Jiaqi Xu, Shenghan Zhou, Hanwen Zhang, Ziheng
  Huang, Chenyu You, and Xiaohui Xie.
\newblock Coma: Compositional human motion generation with multi-modal agents,
  2025.

\bibitem[Taylor et~al.(2021)Taylor, Windle, Greenwood, and Matthews]{Taylor}
Sarah~L. Taylor, John~T. Windle, David Greenwood, and Iain Matthews.
\newblock Speech-driven conversational agents using conditional flow-vaes.
\newblock \emph{Proceedings of the 18th ACM SIGGRAPH European Conference on
  Visual Media Production}, 2021.

\bibitem[Tevet et~al.(2022)Tevet, Gordon, Hertz, Bermano, and
  Cohen-Or]{motionclip}
Guy Tevet, Brian Gordon, Amir Hertz, Amit~H Bermano, and Daniel Cohen-Or.
\newblock Motionclip: Exposing human motion generation to clip space.
\newblock In \emph{European Conference on Computer Vision}, pages 358--374.
  Springer, 2022.

\bibitem[Tian et~al.(2023)Tian, Zhang, Liu, and Wang]{recoverysurvey}
Yating Tian, Hongwen Zhang, Yebin Liu, and Limin Wang.
\newblock Recovering 3d human mesh from monocular images: A survey.
\newblock \emph{IEEE Transactions on Pattern Analysis and Machine
  Intelligence}, 45\penalty0 (12):\penalty0 15406--15425, 2023.

\bibitem[Van Den~Oord et~al.(2017)Van Den~Oord, Vinyals, et~al.]{VQ-VAE}
Aaron Van Den~Oord, Oriol Vinyals, et~al.
\newblock Neural discrete representation learning.
\newblock \emph{Advances in neural information processing systems}, 30, 2017.

\bibitem[Vaswani et~al.(2017)Vaswani, Shazeer, Parmar, Uszkoreit, Jones, Gomez,
  Kaiser, and Polosukhin]{Transformer}
Ashish Vaswani, Noam Shazeer, Niki Parmar, Jakob Uszkoreit, Llion Jones,
  Aidan~N Gomez, \L~ukasz Kaiser, and Illia Polosukhin.
\newblock Attention is all you need.
\newblock In \emph{Advances in Neural Information Processing Systems}. Curran
  Associates, Inc., 2017.

\bibitem[Wagner et~al.(2014)Wagner, Malisz, and Kopp]{RuleBaseReview3}
Petra Wagner, Zofia Malisz, and Stefan Kopp.
\newblock Gesture and speech in interaction: An overview.
\newblock \emph{Speech Commun.}, 57:\penalty0 209--232, 2014.

\bibitem[Yao et~al.(2024)Yao, Zhang, Sun, and Tang]{recovery1}
Wei Yao, Hongwen Zhang, Yunlian Sun, and Jinhui Tang.
\newblock Staf: 3d human mesh recovery from video with spatio-temporal
  alignment fusion.
\newblock \emph{IEEE Transactions on Circuits and Systems for Video
  Technology}, pages 1--1, 2024.

\bibitem[Yi et~al.(2023)Yi, Liang, Liu, Cao, Wen, Bolkart, Tao, and
  Black]{talkshow}
Hongwei Yi, Hualin Liang, Yifei Liu, Qiong Cao, Yandong Wen, Timo Bolkart,
  Dacheng Tao, and Michael~J Black.
\newblock Generating holistic 3d human motion from speech.
\newblock In \emph{Proceedings of the IEEE/CVF Conference on Computer Vision
  and Pattern Recognition}, pages 469--480, 2023.

\bibitem[Yoon et~al.(2020)Yoon, Cha, Lee, Jang, Lee, Kim, and Lee]{FGD}
Youngwoo Yoon, Bok Cha, Joo-Haeng Lee, Minsu Jang, Jaeyeon Lee, Jaehong Kim,
  and Geehyuk Lee.
\newblock Speech gesture generation from the trimodal context of text, audio,
  and speaker identity.
\newblock \emph{ACM Transactions on Graphics (TOG)}, 39\penalty0 (6):\penalty0
  1--16, 2020.

\bibitem[Zhang et~al.(2021)Zhang, Tian, Zhou, Ouyang, Liu, Wang, and
  Sun]{recovery3}
Hongwen Zhang, Yating Tian, Xinchi Zhou, Wanli Ouyang, Yebin Liu, Limin Wang,
  and Zhenan Sun.
\newblock Pymaf: 3d human pose and shape regression with pyramidal mesh
  alignment feedback loop.
\newblock In \emph{Proceedings of the IEEE/CVF International Conference on
  Computer Vision (ICCV)}, pages 11446--11456, 2021.

\bibitem[Zhang et~al.(2023{\natexlab{a}})Zhang, Tian, Zhang, Li, An, Sun, and
  Liu]{recovery2}
Hongwen Zhang, Yating Tian, Yuxiang Zhang, Mengcheng Li, Liang An, Zhenan Sun,
  and Yebin Liu.
\newblock Pymaf-x: Towards well-aligned full-body model regression from
  monocular images.
\newblock \emph{IEEE Transactions on Pattern Analysis and Machine
  Intelligence}, 45\penalty0 (10):\penalty0 12287--12303, 2023{\natexlab{a}}.

\bibitem[Zhang et~al.(2023{\natexlab{b}})Zhang, Zhang, Cun, Zhang, Zhao, Lu,
  Shen, and Shan]{t2m-gpt}
Jianrong Zhang, Yangsong Zhang, Xiaodong Cun, Yong Zhang, Hongwei Zhao, Hongtao
  Lu, Xi Shen, and Ying Shan.
\newblock Generating human motion from textual descriptions with discrete
  representations.
\newblock In \emph{Proceedings of the IEEE/CVF conference on computer vision
  and pattern recognition}, pages 14730--14740, 2023{\natexlab{b}}.

\bibitem[Zhang et~al.(2024)Zhang, Ao, Zhang, Gao, Lin, Chen, and
  Liu]{SemanticGS}
Zeyi Zhang, Tenglong Ao, Yuyao Zhang, Qingzhe Gao, Chuan Lin, Baoquan Chen, and
  Libin Liu.
\newblock Semantic gesticulator: Semantics-aware co-speech gesture synthesis.
\newblock \emph{ArXiv}, abs/2405.09814, 2024.

\bibitem[Zhou et~al.(2022)Zhou, Yang, Li, Saito, Aneja, and Kalogerakis]{Zhou}
Yang Zhou, Jimei Yang, Dingzeyu Li, Jun Saito, Deepali Aneja, and Evangelos
  Kalogerakis.
\newblock Audio-driven neural gesture reenactment with video motion graphs.
\newblock \emph{2022 IEEE/CVF Conference on Computer Vision and Pattern
  Recognition (CVPR)}, pages 3408--3418, 2022.

\bibitem[Zhu et~al.(2023)Zhu, Liu, Liu, Qian, Liu, and Yu]{diffgesture}
Lingting Zhu, Xian Liu, Xuanyu Liu, Rui Qian, Ziwei Liu, and Lequan Yu.
\newblock Taming diffusion models for audio-driven co-speech gesture
  generation.
\newblock In \emph{Proceedings of the IEEE/CVF Conference on Computer Vision
  and Pattern Recognition (CVPR)}, pages 10544--10553, 2023.

\end{thebibliography}
}

\end{document}